\documentclass{cmspaper}
\begin{document}


\begin{titlepage}

   \cmsnote{2000/066}
   \date{\today}

  \title{Application of Neural Networks for Energy Reconstruction}

  \begin{Authlist}
    J.~Damgov\Aref{a} and L.~Litov
       \Instfoot{cern}{Sofia University,Sofia, Bulgaria}
  \end{Authlist}


  \Anotfoot{a}{Now at INRNE, Sofia, Bulgaria}

  \begin{abstract}
$\qquad$ The possibility to use Neural Networks for reconstruction of the
energy deposited in the calorimetry system of the CMS detector is
investigated. It is shown that using feed --- forward neural network, 
good
linearity, Gaussian energy distribution and good energy resolution can be
achieved.   Significant improvement of the energy resolution and
linearity is reached in comparison with other weighting methods for
energy reconstruction. 
  \end{abstract}
\submitted{Submitted to {\it Nuclear Instruments and Methods A}}

\end{titlepage}

\setcounter{page}{2}

\section{Introduction}

$\qquad$ Compact Muon Solenoid (CMS) \cite{CMS} is one of the two
general-purpose major detectors for the LHC accelerator. The main goal of
this experiment is a precise test of the Standard Model and observation of
the Higgs boson as well as the search for diverse signatures of new
physics. It will be achieved  by identifying and precisely measuring
muons, electrons and photons over a large energy range, by determining the
signatures of quarks and gluons through the measurement of jets of charged
and neutral particles (hadrons) with moderate precision, and by measuring
the missing transverse energy flow, which will enable the signatures of
non-interacting new particles as well as neutrinos to be identified.

$\qquad$ The Standard Model Higgs boson with a mass between 95 and 150
$GeV$ would be discovered via its two-photon decay, with a mass between
135
and 525 $GeV$ --- in the four-lepton channel. Tagging the events produced
by WW- and ZZ-fusion by detecting the characteristic forward jets, and
using decay modes with larger branching ratios (H $\longrightarrow$ WW
$\longrightarrow$ lvjj, and H $\longrightarrow$ ZZ $\longrightarrow$
lljj) should allow the discovery range for SM Higgs boson to be extended
up to 1 TeV .  The two-photon and four-lepton channels are also crucial
for the discovery of the Higgs boson in the Minimal Supersymmetric
Standard Model.  Events with many high energy jets and large missing
transverse energy are the obvious and model independent signature in
searches for the supersymmetric partners of quarks and gluons. 

$\qquad$The CMS calorimeters will play a significant role in exploiting
the physics potential offered by LHC. The main functions are to identify
and measure precisely the energy of photons and electrons, to measure the
energy of jets, and to provide hermetic coverage for measuring the missing
transverse energy. In addition, good efficiency for electron and photon
identification as well as excellent background rejection against hadrons
and jets are required.  

$\qquad$ One of the principal CMS objectives is to construct a very high
performance electromagnetic calorimeter (ECAL). A homogeneous scintillating
crystal calorimeter with high granularity gives the best performance for
energy resolution enhancing the H $\longrightarrow \gamma\gamma$ discovery
potential at the initially lower luminosities. The choice of crystals for
ECAL, as well as the thinness of the barrel calorimetry, imposes severe
constraints on the hadron calorimeter (HCAL) design and tempers its
performance. In particular, the CMS calorimeter system is strongly
non-compensating. This leads to a non-Gaussian energy distribution and
non-linear response in energy to hadrons.

$\qquad$In order to ensure a Gaussian response and good linearity of the
calorimetry systems, different weighting methods for reconstruction of the
hadron energy, which have the effect of simultaneously minimizing the
non-linearity and the energy resolution can be employed. 

$\qquad$The present paper is devoted to the application of a feed-forward
neural network  for reconstruction of the energy deposited in the
CMS calorimeter system by hadrons and jets. A significant improvement of
the energy resolution and linearity in comparison with other weighting
methods has been achieved.

$\qquad$The paper is organized as follows: In section 2 a brief
description of the CMS and its calorimeter system is presented. In section
3, some aspects of development of hadron showers in the sampling
calorimeters are considered and two of the conventional weighting methods
for reconstruction of the pion and jet energy in the case of CMS are
applied. A basic information about neural networks and their usage for
processing signals from calorimetry systems is given in section 4. In
section 5, details of our approach to the application of neural networks
for energy reconstruction and the particular feed-forward network used in
the case of CMS detector are described. The results obtained are presented
in section 6.

\section{ CMS Calorimeter system }

$\qquad$ The CMS \cite{CMS} detector has an overall length of 21.6 m, with
a calorimeter coverage up to a pseudorapidity of $|\eta |=5$, a radius of
7.5 m, and a total weight of about 12500 t (see figure 1.CMS). CMS
consists of a powerful inner tracking system, a scintillating crystal
electromagnetic calorimeter followed by a sampling hadron calorimeter
mounted inside the cryostat vessel of the 4T Solenoidal Superconducting
Magnet, 13 m long with an inner diameter of 5.9 m. The Magnet is
surrounded by 5 "wheels" (cylindrical structure) and 2 endcaps (disks) of
muon absorber and muon tracking chambers composing the muon detector
system. 

$\qquad$ The CMS tracker consists of a silicon pixel barrel and forward
disks, followed by silicon microstrip devices placed in a barrel and
forward disk configuration.  The tracker is located inside the calorimeter
system and is supported by it. 
  
$\qquad$ The Electromagnetic calorimeter \cite{ECALTDR} comprises a barrel
($|\eta| < 1.479$) and two endcaps.   The active medium is made out of
scintillating lead tungstate crystals ($PbWO_{4}$). The choice is based on
the following properties of these crystals: a short radiation length of
0.89 cm, a small Moliere radius of 2.19 cm, the scintillating process is
fast --- 85 $\%$ of the light is emitted in 20 ns. The transverse
granularity of $\triangle\eta$ x $\triangle\phi$= $0.0175$ x  
$0.0175$. corresponds to a crystal front face of about 22 x 22 $mm^{2}$.
In the endcaps $1.48 <|\eta| < 3.0$, the granularity  increases
progressively to a maximum value of $\triangle\eta$ x $\triangle\phi$=
$0.05$ x $0.05$. The total thickness of about 26 radiation lengths,
corresponding to a crystal length 23 cm is enough to limit the
longitudinal shower leakage to an acceptable level. The presence of a
preshower (3 $X_{0}$ of lead) in the endcap region allows the use of
slightly shorter crystals (22 cm). The light produced by an incident
particle is detected by avalanche photodiodes in the barrel and by vacuum
phototriodes in the endcap. 
  
$\qquad$ The CMS central hadron calorimeter \cite{HCALTDR}, also consists
of barrel (HB) and endcaps (HE) parts, covering the central rapidity range
($|\eta| < 3.0$). Both the barrel and endcap calorimeters experience the 4
Tesla field of the CMS solenoid and hence are necessarily fashioned out of
non-magnetic material (copper alloy and stainless steel). The central
hadron calorimeter is a sampling calorimeter: it consists of active
material inserted between copper absorber plates. The absorber plates are
5 cm thick in the barrel and 8 cm thick in the endcap. The active elements
are 4 mm thick plastic scintillator tiles read out using wavelength
shifting plastic fibres.  The tiles are arranged in a tower structure
pointing to the interaction center.  The lateral segmentation of
$\triangle\eta$ x $\triangle\phi$= $0.087$ x $0.087$ has been chosen so as
not to degrade di-jet mass resolution for highly-boost di-jets. Only in
the region near $\eta = 3$ the segment size is 0.17 x 0.17.
 
$\qquad$ The barrel hadron calorimeter is about 89 cm deep, which at $\eta
= 0$ is only 5.82 nuclear interaction lengths ($\lambda$) in thickness. To
ensure adequate sampling depth for the entire ($|\eta| < 3.0$) region the
first muon absorber layer is instrumented with scintillator tiles to form
an Outer Hadron Calorimeter (HO). The two layers of scintillator of the
barrel HO are divided into the same granularity as the HB and envelope the
entire first layer of the muon iron absorber. In the region $0.  <|\eta| <
0.4$  additional 15 cm thick steel plates are placed in front of muon
chambers. In this region HO consists of 3 sampling layers. In the endcap
region, the HO has only  single sampling layers. 

$\qquad$ To extend the hermeticity of the central hadron calorimeter
system up to $\eta = 5$, in order to have good missing transverse energy
measurement, CMS employs a separate forward calorimeter (HF) located 6 m
downstream of the HE endcaps. The HF calorimeter covers the region $3.0 <
|\eta| < 5.0$. It uses quartz fibres as the active medium, embedded in a
copper absorber matrix.  
  
$\qquad$ In order to minimize non-Gaussian tails in the hadron energy
distribution and to ensure a  linear response to hadron energy, the
inner barrel hadron calorimeter is divided radially (in depth) into two
sampling hadron compartments (HB1 and HB2). There is an initial layer of
sampling immediately following the ECAL electronics, and 17 layers of
sampling  coupled together into single tower read-out. The HE is also
segmented into two different sampling compartments (HE1 and HE2). There
again is initial sampling layer, followed by 18 layers  joined
into a single tower read-out. The HO layers form towers matching the
inner hadron calorimeter granularity and are read out separately. In this
way, for reconstruction of the hadron energy, in every ($\eta, \phi$)
tower , signals from four longitudinal read-outs (ECAL, HB1, HB2 and HO)
are used.

$\qquad$An important issue is the absolute calibration of the
CMS calorimeter system, used to determine the energy scale.
 For the ECAL, the individual calibration of all crystal is
foreseen in two steps. In order to establish an extremely clean set of
high precision initial calibration coefficients for all channels, all
crystals  will be scanned in the CERN SPS test electron beam at two
energies. After installation of the calorimeter in the CMS detector, an
in situ calibration with physics events is envisaged. The most suitable
channels for this purpose will be $Z \longrightarrow e^{-} e^{+}$
which gives energetic correlated electrons in different regions of
the calorimeter. $E/p$ measurements for isolated electrons also is
a high-rate tool especially important at low luminosity.  During the run
of
CMS experiment, a light monitoring system will track the behaviour of each 
channel. 
  
$\qquad$The HCAL calibration and monitoring is designed to determine the
absolute energy scale and monitor the calorimeter system for changes. 
Initially, several individual wedges will be placed in the test beam at
CERN where hadrons and muons of various energies will be used to determine
absolute calibration between beam energy and light yield response to the
moving radioactive wire source \cite{HCALTDR}. The absolute single hadron
test beam calibration will be transferred to the rest of the calorimeter
with the help of the same radioactive source. The dependence of the
calorimeter response on the magnetic field \cite{HCALTEST} makes in situ
calibration using physics events obligatory. In the CMS detector, single
tracked hadrons from $\tau$-leptons, jet balancing ($photon+jet$ and $Z +
jet$) and
di-jet resonances ($W \longrightarrow jj$ in top decays, $Z \longrightarrow
b\bar b$ and $Z \longrightarrow \tau\bar\tau$) will be used for in situ
calibrations.

\begin{center} 
\includegraphics{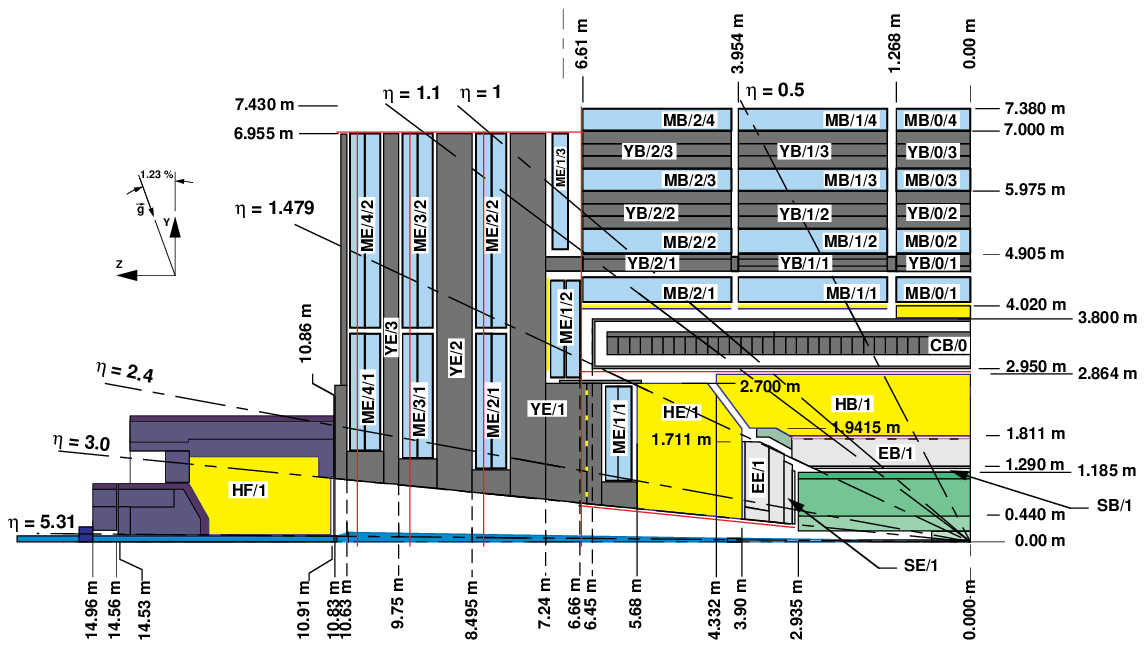}\\ fig.1  CMS 
\end{center} %

\section{Energy reconstruction }

$\qquad$The reconstruction of the energy of a particle entering the
calorimetry system is a nontrivial task. The difficulties are determined ,
first of all, by the complicity of the physical processes taking place in
the development of the electromagnetic and hadron showers and the
mechanism of creation of the signal in the active elements of the
detectors. This leads to  big fluctuations in the signals of the
calorimetry system, especially in the case of hadrons  (the so
called intrinsic fluctuations). In the case of sampling calorimeters, only
a fraction of the shower energy is dissipated in the active medium and the
energy resolution is affected significantly by the fluctuations in this
fraction (sampling fluctuations). Hadron calorimeters , because the large
depth required to incorporate almost the full shower, are by necessity
sampling calorimeters. 

$\qquad$There is an essential difference between development
of a shower initiated by electrons and photons (electromagnetic ) and
the one initiated by hadrons. In the first case (assuming enough deep
calorimeter) high energy electron or photon incident on a calorimeter
initiates a cascade of secondary electrons and photons via bremstrahlung
and pair production. The multiplication continues until the energies of
secondary particles fall bellow the critical energy. The further
dissipation of energy is dominated by ionization and excitation. 
The visible energy is proportional to the energy dissipated in the
calorimeter even in the case of sampling calorimeter.  

$\qquad$The mechanism for hadron shower production is quite different
since it involves multiparticle production in deep inelastic
hadron-nucleus collisions at high energies and thus the elementary
processes are determined by the strong interactions. The large number of
possible interaction processes makes the shower development more complex.
The cascade is mostly composed of nucleons and pions. A significant part
of pions are $\pi^{0}$ ($\sim 30 \%$), as a result the  cascade
contains two distinct components namely the electromagnetic one
($\pi^{0}$'s etc)  and the hadron one ($\pi^{\pm}$, n, etc). The
electromagnetic component is determined essentially by the first
interaction and is subject of considerable event-to-event non-Gaussian
fluctuations. It increases with energy due to the fact that the neutral
pions, developing as electromagnetic showers, do not produce any hadron
interactions. The response of hadron sampling calorimeter is a function
of the energy deposited in the active elements by electromagnetic
component, charged hadrons, low energy neutrons and energy lost in
breaking up nuclei (invisible energy). The last one can be large enough
and for heavy absorbers it reaches up to 40 $\%$ of hadron part of the
energy.  Thus the ratio of response to electromagnetic and hadron
showers ($e/\pi$) is usually $> 1$ and is a function of the initial energy
(non-compensating calorimeters).  This leads to a non-Gaussian measured
distribution for mono-energetic hadrons, a non-linear response in energy
to hadrons and an additional contribution to the relative energy
resolution ($\sigma/E$). It is possible to achieve a compensation
($e/\pi=1$) by suppression of the electromagnetic signal or by boosting
the non e.m. signal. However the requirement for precise electromagnetic
calorimetry is non compatible with compensation and leads, as a rule, to
strongly non-compensating calorimetry systems. 

$\qquad$Another problem caused by non-compensation is the dependence of
the calorimeter response on  the type of the incident particle. The
response for single hadrons and jets is different, because significant
part of the jet energy is carried by electrons and photons. Due to the
different cross-sections for creation of $\pi^{0}$-mesons even for pions
and protons the response of the calorimetry system is not equal. The CMS
calorimetry system is a typical example for strongly non-compensating
hadron calorimeter.

$\qquad$ In order to achieve a good linearity, Gaussian energy
distribution and the best possible energy resolution, different methods
for reconstruction of the energy deposited in the calorimetry system are
used. The idea is to find such calibration coefficients (weights) which
 take into account the peculiarities of the development of the hadron
shower. In order to investigate the applicability of the various
reconstruction techniques in the case of CMS calorimetry system we have
simulated the detector response to electrons and pions (with energies 5,
10, 20, 50, 100, 200, 300 and 500 GeV) and single jets (with energies
30,50,100,200,300 and 500 GeV) entering the ECAL at $\eta = 0.1$ and
$\phi=0.1745$.

$\qquad$In the most common approach (in what follows we shall refer to it
as
to SM ), the energy of a single shower is given by the weighted sum of
the energies deposited by it in the calorimetry system
\begin{equation}
E_{rec}=\sum_{j}w_j E_j
\end{equation}
where the index $j=1,2,3,4$ corresponds to the four longitudinal
compartments of the calorimeter (ECAL,HB1, HB2 and HO) and $E_j$ is  
the sum of the energies measured in the transversal towers in the $j'th$
 readout. Given a large enough number of showers with known incident
energy $E_{inc}$, the weights are determined by minimization of the width
of the energy distribution with additional constraint $<E>=E_{inc}$.
   
\begin{figure}[hbtp]
  \begin{center}
    \resizebox{16cm}{!}{\includegraphics{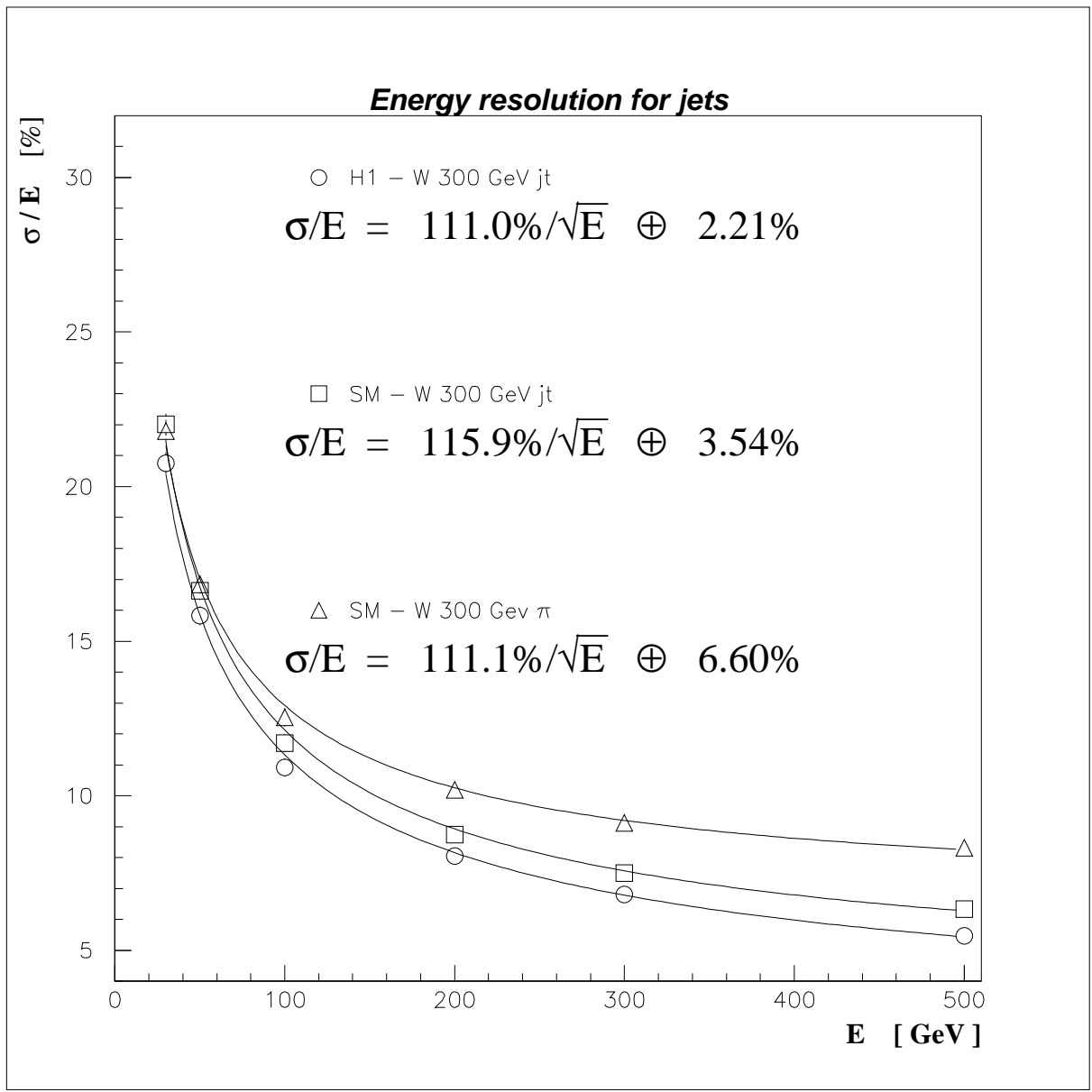}}
    \caption{Energy resolution for jets: Standard Method and H1 technique
with calibration coefficients for 300 GeV.}
    \label{fig:res300}
  \end{center}
\end{figure}

\begin{figure}[hbtp]
  \begin{center}
    \resizebox{16cm}{!}{\includegraphics{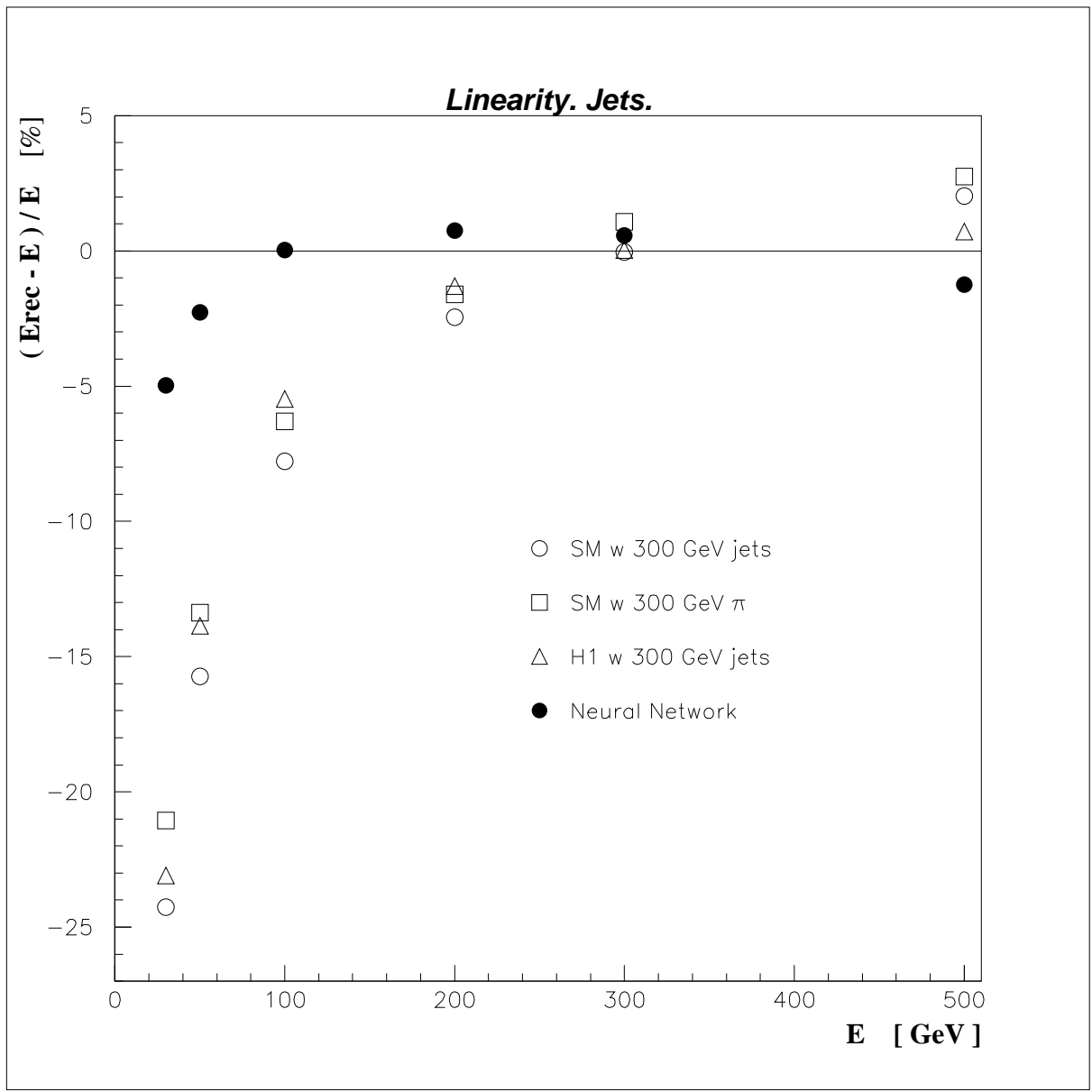}}
    \caption{Linearity for jets}
    \label{fig:lin}
  \end{center}
\end{figure}

\begin{figure}[hbtp]
  \begin{center}
    \resizebox{16cm}{!}{\includegraphics{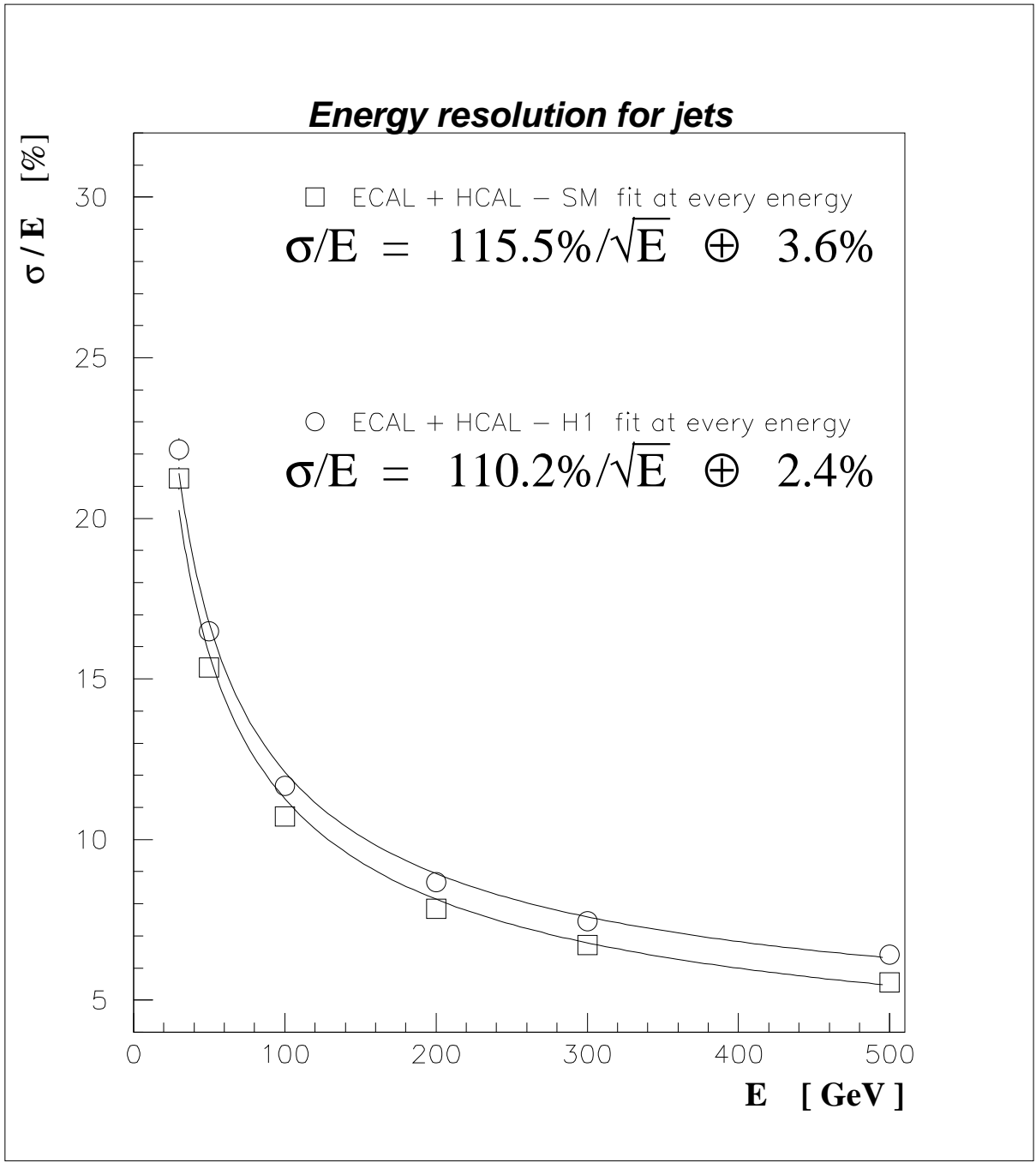}}
    \caption{Energy resolution for jets obtained using  Standard Method
and H1 
technique with $w_i$  fitted at every energy.}
    \label{fig:smh1fitr}
  \end{center}
\end{figure}

$\qquad$We have used the simulated pion data to determine the coefficients
$w_j$. The weights are energy dependent. This dependence is very strong
for $w_1$, which is not surprising, taking into account, that for the ECAL
$e/h=1.7$. If $w_1$ is fixed at the value obtained by using electron
calibration,
 big non-linearity and non-Gaussian tails in the energy distributions for
pions and jets are observed, especially at low energies.  The way to
improve the situation is to determine  $w_1$ using pions, but this
requires a very clear identification in the off-line analysis of the type
of the particles.  The jet energy resolution evaluated by using the
weights
obtained for 300 GeV pions (this corresponds to calibration of the
calorimeter in 300 GeV pion beam, see the previous section)  is
plotted on Fig.~\ref{fig:res300}. The linearity defined as

                   $$ L=\frac{(E_{rec}-E_{inc})}{E_{inc}}$$ 

is shown on Fig.~\ref{fig:lin}. The different electromagnetic content
in the hadron showers developed by pions and jets and the
non-compensation of the calorimetry system lead to the nonlinear response
and induces a big constant term in the energy resolution. We have used the
weights obtained for 300 GeV jets (that corresponds to in situ
calibration with physical events) to reconstruct the energy. As a result
the linearity has been improved slightly (see Fig.~\ref{fig:lin})
and the constant term of the energy resolution has been reduced
significantly (Fig.~\ref{fig:res300}). However this step does not solve
the problem with non-Gaussian tails in the energy distributions and non
linear response of the calorimeter. When we use energy dependent
weights for reconstruction of the energy, the linearity is restored but
the energy resolution does not change ( Fig.~\ref{fig:smh1fitr}).

$\qquad$In the SM method the weights are sensible to the average of  
the fluctuations in the development of the hadron shower. When applied to
individual events, they always involve large errors. A better approach is
to
use a method where a different correction factor is applied to 
each event, depending on its nature. Several weighting techniques
have been developed \cite{H11,H12,H13,H14}  with the
idea  to suppress  the signal from the electromagnetic component of 
hadron shower (the so called H1-technique). The H1-technique we apply is
of the following form: 
\begin{equation}
E_{rec} = \sum_i(w_i\sum_jE_{i,j}-v_{i}\frac{\sum_jE^2_{i,j}}
{\sum_jE_{i,j}}),
\end{equation}
where $E_{i,j}$ is the energy deposited in the j'th transversal tower of
the i'th longitudinal read-out. The weights $w_i$
and $v_i$ have been determined in the same way like in the SM approach.
The
results obtained are plotted on  Fig.~\ref{fig:res300},
Fig.~\ref{fig:lin} and  Fig.~\ref{fig:smh1fitr}. The energy
resolution is slightly improved mainly due to the reduction of the
constant
term. However the nonlinearity is still too big, especially at lower
energies. Obviously within  these methods it is not possible
to achieve good linearity and Gaussian response.

$\qquad$To ensure the best possible measurement of the energy, when
dealing with noncompensating calorimeters, the following requirements
should be satisfied:

$\qquad$--- to every individual event , different correction factor should
be
applied (due to the big fluctuations of the hadron shower development);

$\qquad$--- using the lateral and longitudinal energy distribution
(the only
available information from the calorimeter), the amount of the  energy
dissipated in the calorimeter by electromagnetic showers
(electromagnetic part of the hadron shower) should be estimated;

$\qquad$--- the type of the particle (electron/photon, hadron or jet)
should
be determined.

$\qquad$Of course the spatial energy distribution does not gives the full
information about the development of the hadron shower, however it is
possible taking into account correlations between signals to determine the
type of the initiator of the shower, and to estimate roughly the energy of
the electromagnetic part of the shower. For example, using the
longitudinal distribution of the energy, it is possible to separate
electrons (big signal in the ECAL and almost nothing in the HCAL) from
pions and jets.  There is a significant difference in the lateral radius
of the electromagnetic and hadron showers --- the radius of the hadron
shower is much bigger then the electromagnetic one. This feature can be
used for estimation of the electromagnetic content of the shower. In some
sense the H1-technique accounts for this difference and manifests better
energy resolution and linearity.

$\qquad$To solve the problem of energy reconstruction we need a method
which is able to deal with many parameters, is sensitive to correlations
between them
and is  flexible enough to react to  fluctuations in the
development of
the hadron shower. One possible solution is the application of neural
networks, which have proved their efficiency  in  such
a complicated environment.

\section{ The neural network approach}

$\qquad$ The neural networks (NN) are a powerful tool, which can be used
for feature extraction, association, optimization, function fitting, and
modeling. They have found numerous applications
\cite{OBZ,DENBY} in the
 High Energy Physics, such as classification of particles and final
states, track reconstruction, cluster trigger, particle identification,
reconstruction of invariant masses and  other off-line analysis  
\cite{NN1,NN2,NN3,NN4,NN5,NN6}.

$\qquad$ The neural network is a system composed of many simple processing
elements operating in parallel whose function is determined by network
structure, connection strengths, and the processing performed at computing
elements or nodes. The basic computing unit of NN is a neuron. A
Multi-Layer Feed Forward (MLFF) network consists of a set of input
neurons,
one or more layers of hidden neurons, and a set of output neurons. The
neurons of each layer are connected to the ones in the subsequent layer. 

$\qquad$The neurons (Fig. 1.NN) perform calculations in three steps via
their input I, activation A and output O functions. Usually  they are
chosen in the following form: 
\begin{equation}
I_i=\sum_{k} w_{ik}O_k, 
\qquad   A_i(I)=\frac{1}{1+e^{-(I_{i} + b_{i})}} , 
\qquad O_i=\Theta(A_{i}-A_{0i}),
\end{equation}
where $w_{ik}$ are the connections weights,  $A_{0i}$ is  some threshold.
In
most of the architectures a bias
$b_{i}$ 
is implemented as an incoming link connected to a constant value  1. The
weight of the connection is trained like all other weights. %

\begin{center} 
\includegraphics{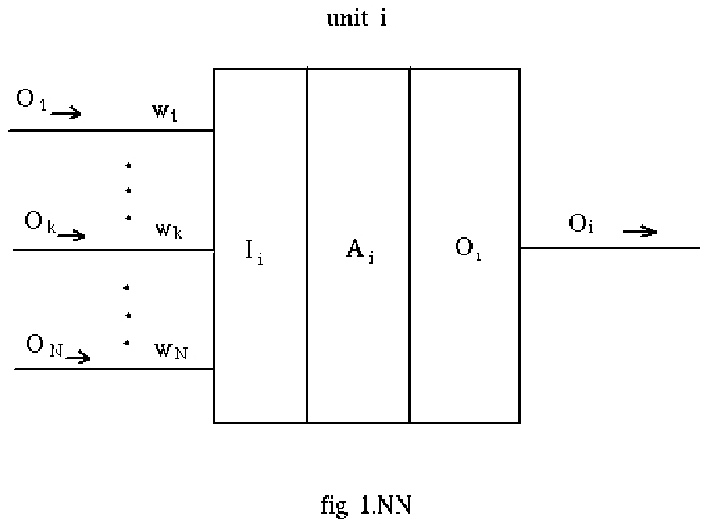} 
\end{center}%

$\qquad$ MLFF networks are, usually, trained using the Backpropagation
learning algorithm \cite{BPNN,NNHEP}. It includes three steps:  
presentation of a pattern to the network,  comparison of the
desired output with the actual network output,  backwards
calculation of the error and adjustment of the weights (wij). This is
done by minimization of the error function,
\begin{equation}
E=\frac{1}{2} \sum_{j} (t_j-o_j)^2 ,
\end{equation}
where $o_j$ is the output of the net and $t_j$ is the corresponding
element of teaching sample. Backpropagation algorithm uses an updating
rule: 

        $$ \Delta w = -\eta  \frac{\partial{E}}{\partial{w}}$$,

where $\eta$ is the learning rate. The optimal value for $\eta$ varies
significantly during the learning of the NN. For that reason we have
chosen another learning algorithm --- Rprop \cite{RPROP}. It uses an
individual learning rate for each weight combined with the Manhattan
updating rule \cite{MHLRNN}: 

           $$\Delta w = - \eta sign [\frac{\partial{E}}{\partial{w}}]$$ 

At every step, $\eta$  is adjusted as: 

$$\eta_{w,t+1}=
   \gamma^+\eta_{w,t} \qquad\mbox{if\qquad 
$\partial{E_{t+1}}.\partial{E_t} > 0
$},$$

$$\eta_{w,t+1}=
    \gamma^-\eta_{w,t} \qquad\mbox{if\qquad 
$\partial{E_{t+1}}.\partial{E_t} <
0 $}$$

$$0<\gamma^-<1<\gamma^+$$

$\qquad$ There are two possible approaches to the problem of the energy
reconstruction with the assistance of NN. The first one is to use NN
directly
to determine the energy dissipated in the calorimetry system. Such an 
approach has been applied in the case of gamma-ray energy determination
with GILDA imaging silicon calorimeter \cite{GILDA}. The energy
reconstruction is performed in two steps. In the first stage, a net
performs a rough classification of the gamma energies in six groups. Then,
for each group, a dedicated net proceeds to discriminate among the
different energy values. At the second stage each net has an output layer,
which
consists of Ni nodes, equal to the number of the different energy
subclasses used in the training phase. The net has a discrete output and
is able to classify only those energies used during the training.  Since
gamma rays have a continuous energy distribution, the following weighted
average over the values of output neurons is performed: 
 
               $$E_\gamma=\frac{\sum_{i}(o_i . e_i)}{\sum_i(o_i)}$$

where $e_i$ is the energy corresponding to gamma rays of class $i$ and
$o_i$ is the activation value of the output neuron $i$. 

$\qquad$ An analogues algorithm for reconstruction of the energy in the
case of CMS detector has been applied.  This scheme has shown significant
distortion of spectrum shape, bad energy resolution and a big
nonlinearity. Obviously, this is a consequence of the more complicated
structure and bigger fluctuations of hadron showers in this case. 

$\qquad$A slightly different approach was used to determine the energy
correction factors for the ATLAS detector \cite{ATLAS} using recurrent
neural network with nearest neighbour feedback in the input layer and a
single output giving the corrected energy value. It was shown that the
network performs satisfactory, but there is no  comparison of the results
obtained in this approach with results given by more conventional
algorithms. 

$\qquad$The second possibility is to use the neural network for adjustment
of the weights $w_{j}$ in (1)  on event by event basis. We shall developed
it in what follows.

\section{Energy reconstruction with Neural Network}

$\qquad$The idea of our method is : to tune the weights (calibration
coefficients) $w_j$ in (1) for the four longitudinal readouts of the
calorimetry system on event by event basis, using the available
information about lateral and transversal development of the hadron
shower. Such an approach is more flexible and allows to take into account
the quite different behaviour (for example --- different e/h-ratio)  of
the electromagnetic and hadron compartments of the CMS calorimetry system.
 
$\qquad$As it was mentioned above, the calorimeter system of CMS is
non-compensating. As a result, there is difference between
optimal calibration coefficients for jets, hadrons and electrons. 
Therefore, in order to improve the energy reconstruction, it is preferable
to identify the type of the initiator of the shower.

$\qquad$ The data processing  is divided in two steps: first, 
identification of the type of the incident particle and second, 
determination of the energy deposited in the calorimeter. Correspondingly,
we  use two-level neural network. The first level network 
classifies the hadron showers in four classes, respectively, initiated
by:
\begin{itemize}
\item mainly electromagnetic interacting particles - $e^+$, $e^-$ and $\gamma$.
\item mainly strong interacting particles - hadrons.
\item jets.
\item muons.
\end{itemize}
                                                       
$\qquad$The energy reconstruction is then performed by a dedicated network
for each class of showers.  The second level network has four
subnets (Fig2.NN) corresponding to the four longitudinal
read-outs. The activation values of the four output neurons give the 
relevant  correction factors for the weights in the SM. 

$\qquad$The neural network has $30$ input nodes:  
\begin{itemize} 
\item $E_{rec}$ - the energy of the shower reconstructed using the
Standard Method with $w_i$ for 300 GeV
\item $\frac{w_iE_i}{E_{rec}}$, $
i=1,2,3,4 $, where  $E_i$ is the energy deposited in the $i$th
read-out.  
\item $13$ inputs with signals from ECAL 
\item $3$ x $4$ inputs with signals from the three read-outs of HCAL 
\end{itemize}
\begin{center} \includegraphics{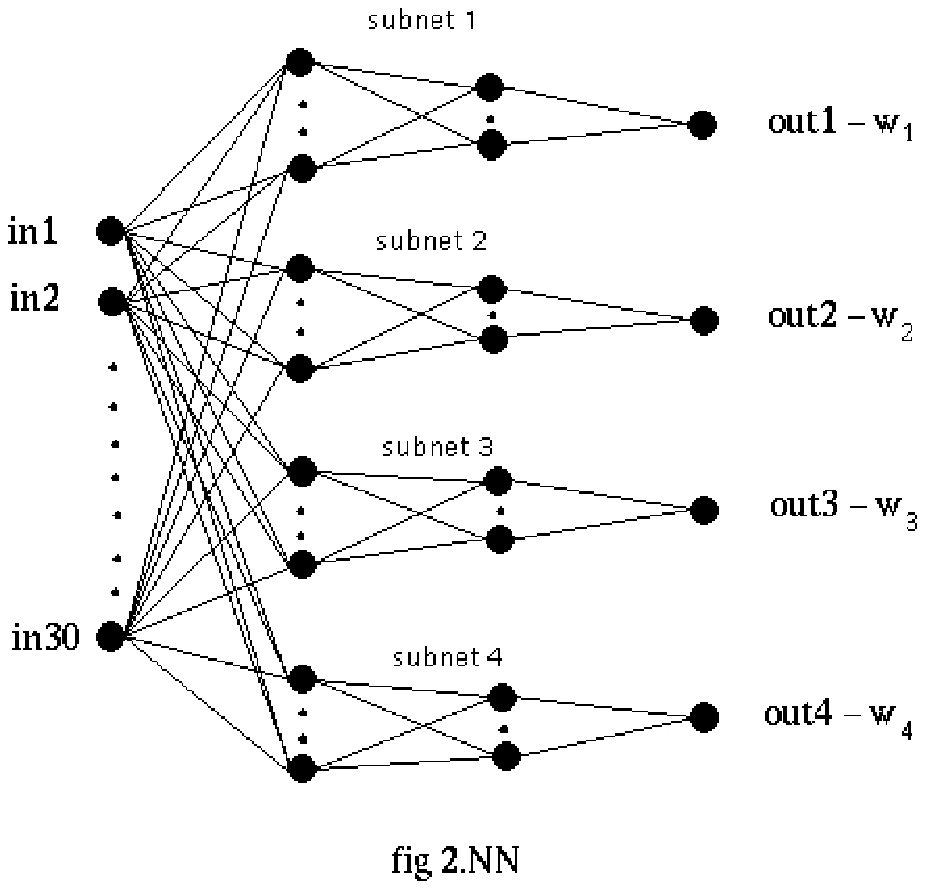} \end{center} %

$\qquad$In our reconstruction scheme, we take into account the energies
deposited in a ($\eta, \phi$) cone with transversal radius $\Delta R =
0.85$ around the shower maximum. This corresponds to matrix of towers with
size 41x41 for the ECAL and with size 7x7 for the HCAL.  Because the
number of towers is too large, in order to reduce the number of network
input neurons, we merge together transversal towers, by summing the
energies in concentric squares as shown on Fig. 3.NN. 

 \begin{center} \includegraphics{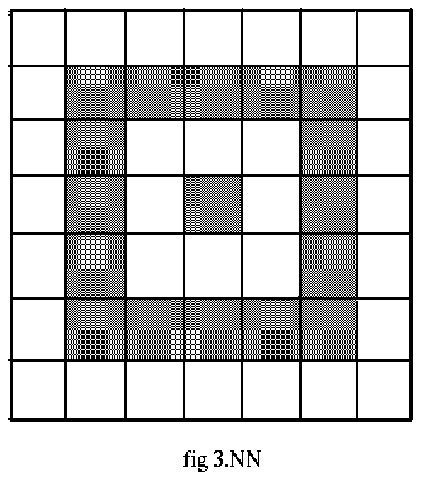} \end{center} %

$\qquad$ During the training phase the NN stands in need of some
additional neurons.  The supplementary part of the network is represented
on Fig. 4.NN .  Outputs
 (out1 ... out4) of the basic network are connected with four hidden
neurons with input function $I(O)=\ O$ and activation function $A(I)=I$
. Four additional inputs (extra-in1 ... extra-in4) multiply them with a 
value, proportional to the energy deposited  in the corresponding read-out
of
the calorimeter.  The output neuron (out)  sums up the signals (A(I)=I). 
Additionally, we correct the output teaching samples with factor
$\frac{1}{\sqrt{(E_{in})}}$ in order to obtain correct contribution from
different energies to the error function (4) of the net. Without this
correction, the contribution to the error function of events with low
energies is minor and as a result, the connection weights of the NN are
tuned to reconstruct well only high energy showers. The same correction is
applied to the values of the extra inputs. During the learning phase, the
weights of the connections ($out_{i}-hid_{i}$) - $w_{i}$,
($extra-in_{i}-hid_{i}$)-$u_{i}$ and ($hid_{i}-out$) - $v_{i}$, are
treated like all other weights in the NN.  After the learning is
finalized, the supplementary part of the net is removed.  

\begin{center}
\includegraphics{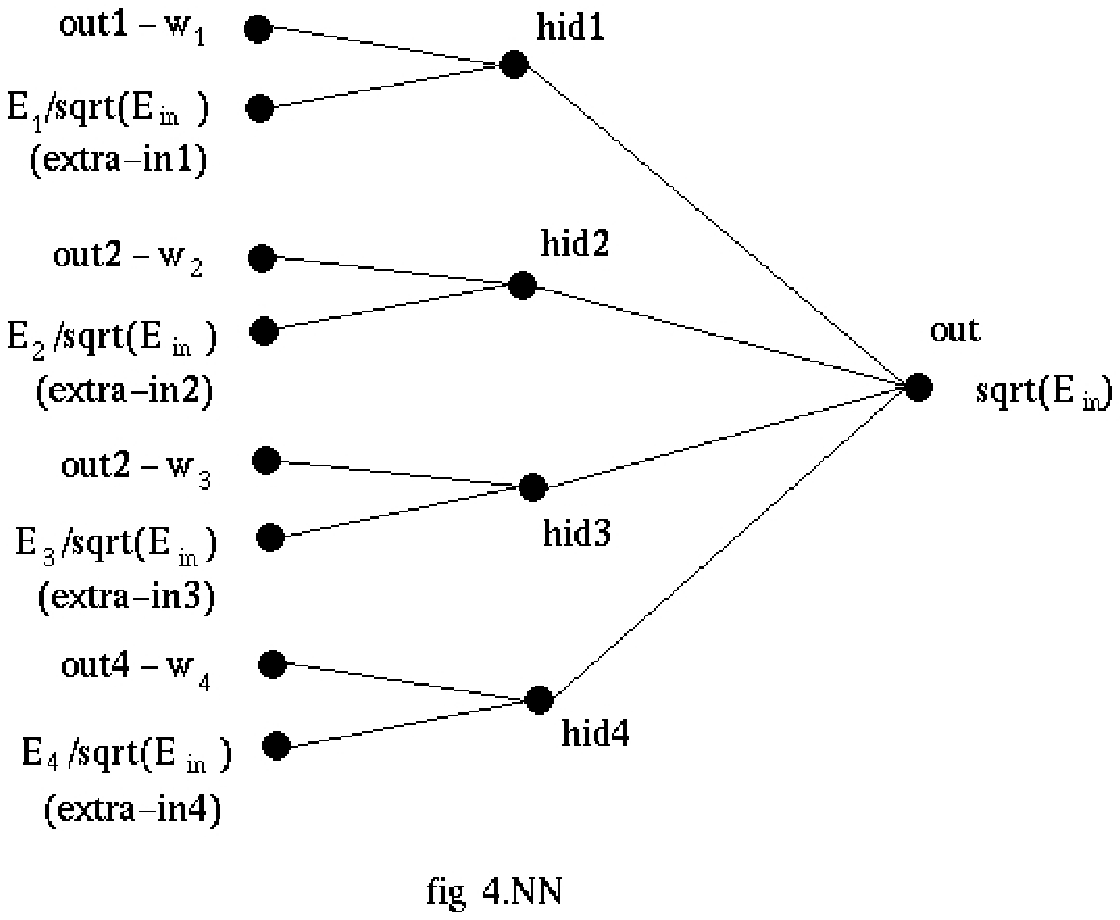}
\end{center} %

$\qquad$The reconstructed energy is calculated as:
\begin{equation}
          E_{rec}= \sum_i o_i.W_i.E_i 
\end{equation}
 where $o_i$ are  activation values of the output neurons of NN, $W_i =
w_i.v_i.u_i$ and $E_i$ are the energies deposited  in the longitudinal
read-outs .The coefficients $W_i$ are fixed during the learning of the NN. 
Variations in the shower development are accounted by $o_i$.

$\qquad$For realization of the neural network we have used Stuttgart
Neural Network Simulator (SNNS) \cite{SNNS}. Our experience has shown that
this simulator is more flexible and efficient, than the widely used in
the high energy physics neural network package JETNET \cite{JN}.

\section{Results}

$\qquad$To test the performance of the neural network we have used the
same sample of simulated events as in section 3. The number of particles
of each energy and type is 5000, 3000 of them have been used to train the
network and the rest 2000 for the test itself. Additional samples of
simulated events with energies different from those used during learning 
phase, have been used as well  to carry out an independent test of the 
network.

$\qquad$ The recognition of the type of the shower initiator, has been
done by two different methods. In the first method,  using suitably
chosen cuts, we separate consequently electrons from pions and
jets and after that, pions from jets.  The first cut we apply is the
value of the shower pseudoradius  in the electromagnetic calorimeter:

$$R_{sh}=\sqrt{\frac{\sum_{i,j}E_{ij}\phi^2_i+\sum_{i,j}E_{ij}\eta^2_j}{\sum_{i,j}E_{ij}}-
\frac{(\sum_{i,j}E_{ij}\phi_i)^2+(\sum_{i,j}E_{ij}\eta_j)^2}{(\sum_{i,j}E_{ij})^2}}$$

where $E_{ij}$ is the energy deposited in the crystal $ij$ with
coordinates $\phi_i$,$\eta_j$. If $R_{sh}$ has value in the interval
(0.25,0.7), the shower is considered as initiated by electron. The
efficiency for recognition of electromagnetic showers is 99.95 $\%$. 

$\qquad$ The second step is separation of single hadron showers.  For
their recognition  several additional cuts have been used. The showers
are classified as induced by single hadron if:  
\begin{itemize} 
\item $R_{sh}<0.07$ 
\item the energy deposited  in the electromagnetic calorimeter corresponds
to
MIP ($0.7 GeV$) 
 \item $mR > 0.332$, where
$mR=\frac{\sum_{i,j}E^{1.5}_{ij}}{\sum_{i,j}E^{1.63}_{ij}}$.  \item
$R_2>37.5$, $R_2=\frac{E_{HCAL}}{E_{ECAL}}$, where $E_{ECAL}$ and
$E_{HCAL}$ are the energies dissipated in the electromagnetic and hadron
calorimeters correspondently. The energies are obtained using calibration
coefficients for 300 GeV pions. 
\end{itemize} 

$\qquad$ If a shower does not
obey any of the above conditions, it is classified as  initiated by hadron
jet. 

$\qquad$ The shower is classified as initiated by muon, if in all
calorimeter compartments , the signal corresponds to the MIP one. This
simple requirement gives very high (more $99 \%$) efficiency of muon
recognition.

\begin{figure}[hbtp]
  \begin{center}
    \resizebox{16cm}{!}{\includegraphics{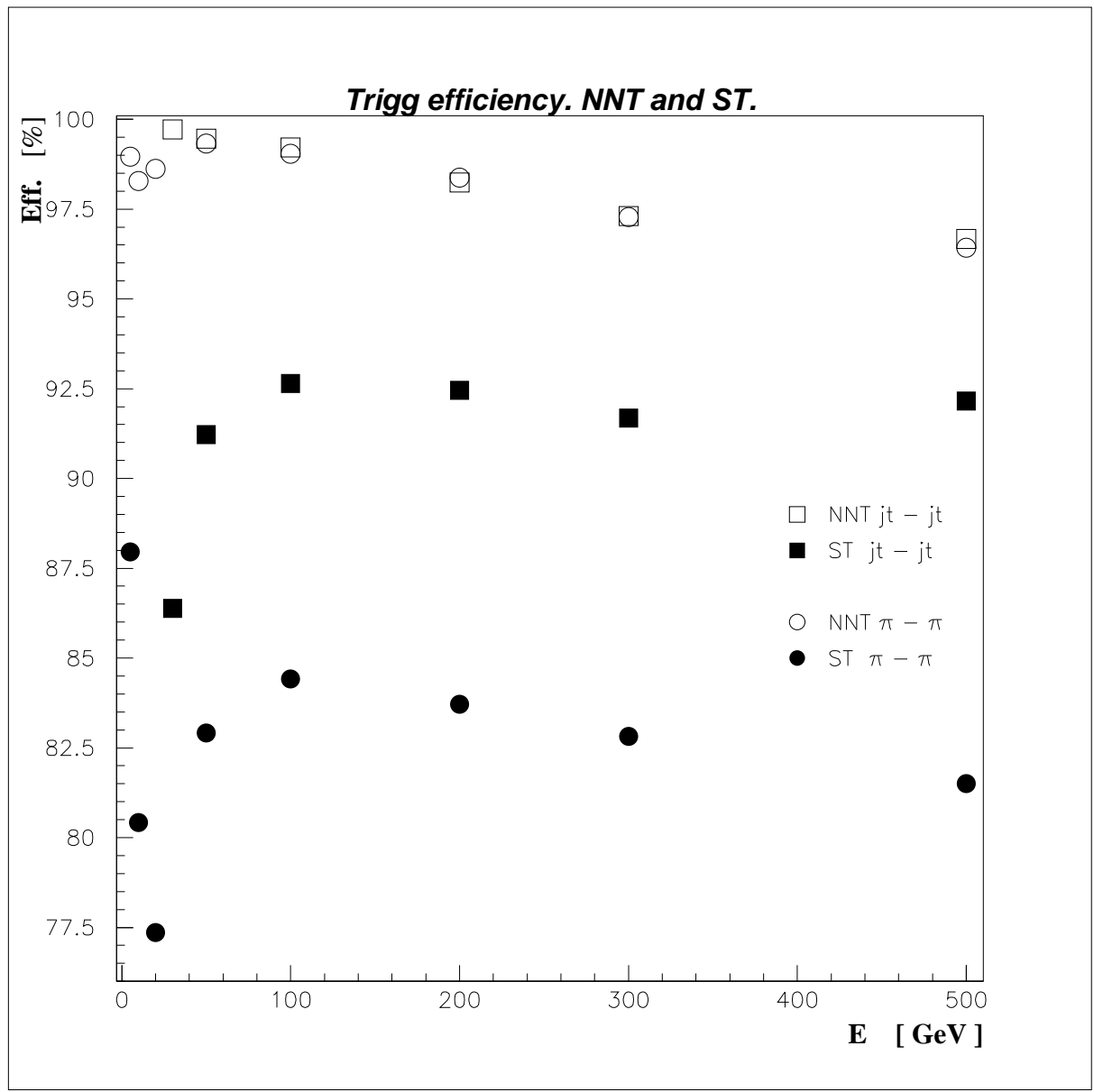}}
    \caption{Efficiency of recognition of the type of the shower
initializing particle. The solid circles and squares are data obtained
using conventional cuts, the empty one represent results obtained using
neural network.}
    \label{fig:nntef}
  \end{center}
\end{figure}

$\qquad$ The separation of electrons from jets and pions is 
efficient enough. However, the efficiency for pion recognition varies with  
the energy between $88\%$ and $93\%$. The situation with jets
is even worse --- the efficiency is between $77\%$ and $84\%$ (see
Fig.~\ref{fig:nntef}). 

$\qquad$In order to get better efficiency of recognition we have used   a
neural network. As we can expect, the efficiency of recognition
is much higher and even for pions and jets is more then
$97\%$ (Fig.~\ref{fig:nntef}).

$\qquad$ Of course it is possible to reach better efficiency of
recognition using
additional information from the central tracker, but for our purposes
it is more important to classify the showers according to their specific
energy distribution in the calorimetry system.
Then  even in the case of wrong recognition of the initializing particle,
it is possible to reconstruct the energy of the shower with good
accuracy.
 The reconstructed energy by NN for 300 GeV jet showers , classified by NN
as
single hadron showers is shown on Fig.~\ref{fig:nnjp}. The reason for
misidentification is that in those cases the shower characteristics are
very close to the one for pion showers. This explains why is nevertheless
the energy sufficiently well reconstructed.  The same situation takes
place for jets mixed with electrons.  The part of wrongly classified
events is small and the reconstruction of their energies is satisfactory.  
This can be seen on Fig.~\ref{fig:nnwrong}, where the
reconstructed energy for 300 GeV jets for all events is plotted and the
contribution of the wrongly recognized events is shown. 

\begin{figure}[hbtp]
  \begin{center}
    \resizebox{11cm}{!}{\includegraphics{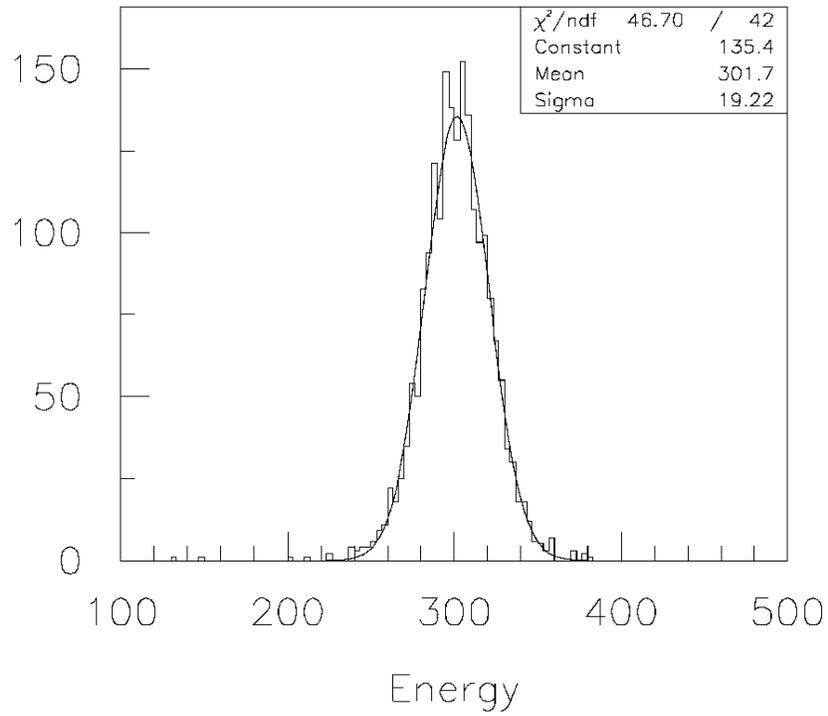}}
    \caption{ Reconstructed energy for 300 GeV jets with NN.}
    \label{fig:nnjs}
  \end{center}
\end{figure}

\begin{figure}[hbtp]
  \begin{center}
    \resizebox{16cm}{!}{\includegraphics{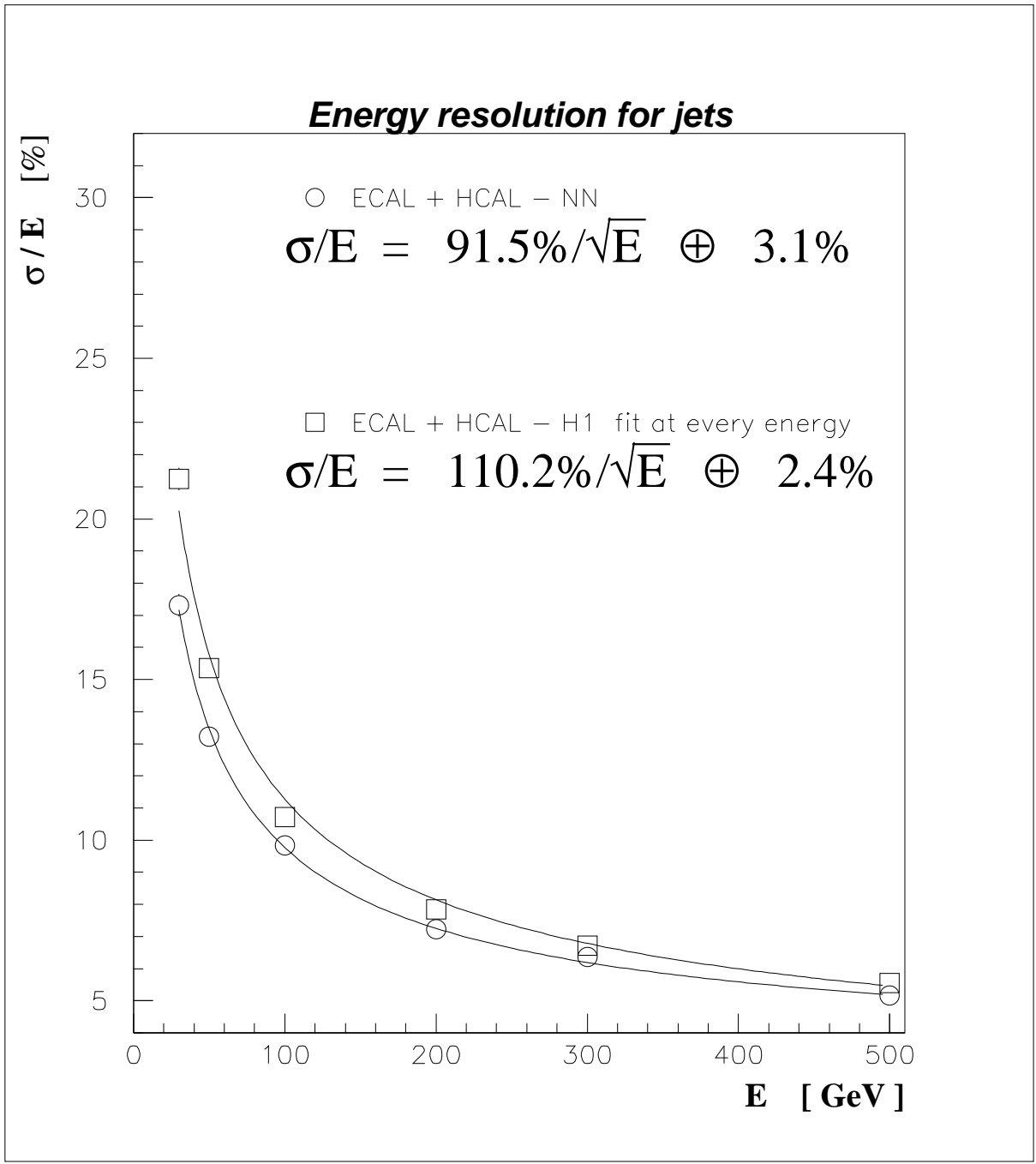}}
    \caption{Energy resolution for jets. Comparison between NN and H1 with
$w_i$  fitted at every energy.}
    \label{fig:nnh1r}
  \end{center}
\end{figure}

$\qquad$ The spectra of reconstructed energies for 300 GeV jets by NN have
clear Gaussian shape (see Fig.~\ref{fig:nnjs}) and even at low energies
there is no significant tails. The energy resolution obtained is
 
$$\frac{\sigma}{E}
=\frac{91.5\%}{\sqrt{E}} \oplus 3.1\%
$$  

and is plotted on Fig.~\ref{fig:nnh1r} in comparison with the one obtained
with the use of the H1 technique with weights fitted at every energy
(idealized case).  The stochastic term is lower than this evaluated with
the H1 technique ( $\frac{\sigma}{E} =\frac{110.2\%}{\sqrt{E}} \oplus
2.4\%$ )  mainly due to the improvement of the resolution for low
energies.  The constant term is slightly higher for NN-based solution. 
Linearity is shown on Fig.~\ref{fig:lin}.  It can be improved further
adding a small NN to the NN performing the energy reconstruction.

\section{Conclusions}

$\qquad$A feed-forward neural network has been applied for reconstruction
of the energy deposited in the CMS detector by single hadrons and jets. We
 perform the reconstruction in two steps. First, the showers are
classified by the NN, according to the type of the initializing particle
(muon, electron, single hadron, hadron jet), A high enough efficiency of
recognition has been achieved. It is shown, that even if the shower is
misidentified, its energy is reconstructed correctly. At the second step,
for the every class of events, a dedicated Neural Network evaluates the
energy of the shower. A significant improvement of the energy resolution
and linearity has been achieved in comparison with the ones acquired with
the help of different weighting methods. The energy spectra have a
Gaussian shape and are free of tails.

\section*{Acknowledgements}We would like to tank V. Genchev and N.
Durmenov for numerous  helpful discussions.

\begin{2figures}{hbtp}
  \resizebox{\linewidth}{1.0\linewidth}{\includegraphics{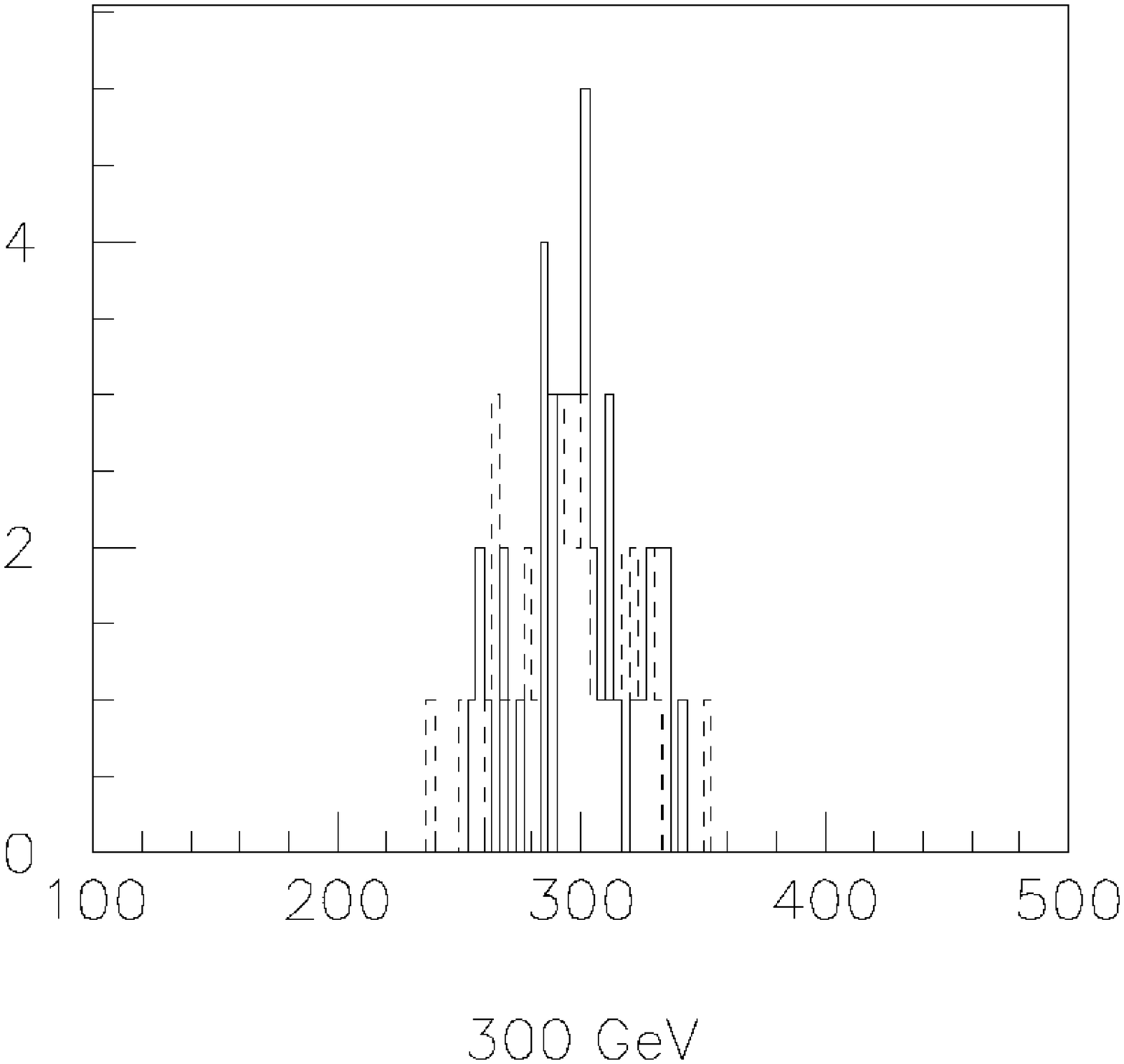}} &
  \resizebox{\linewidth}{1.0\linewidth}{\includegraphics{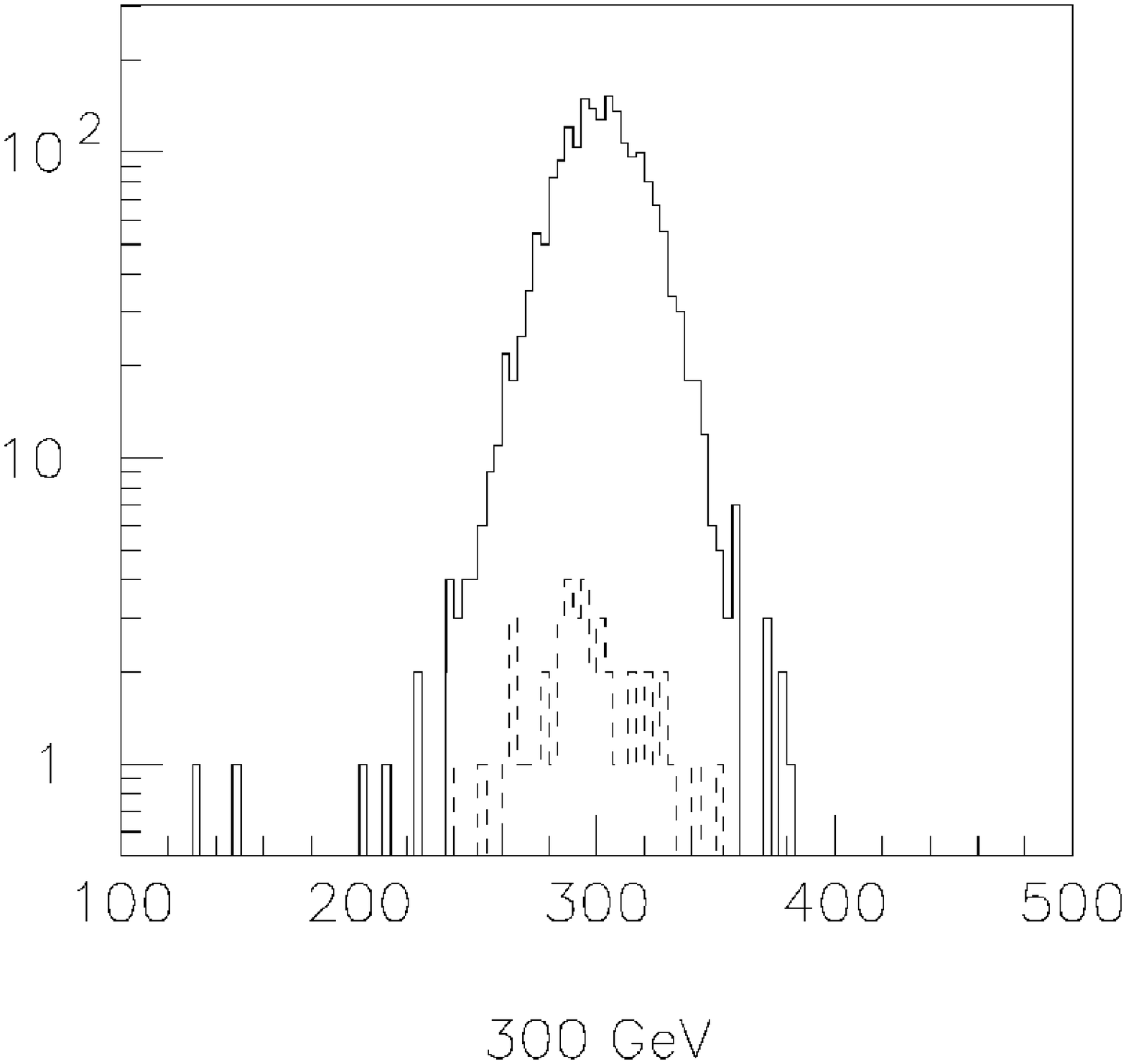}} \\
  \caption{Reconstructed energy with NN for 300 GeV jets misidentified as
pions. Dash line - reconstructed as pions, solid line - as jets.}
  \label{fig:nnjp} &
  \caption{Reconstructed energy with NN for 300 GeV jets (solid line). 
Dashed line - contribution of misidentified jets.}
  \label{fig:nnwrong} \\
\end{2figures}

\end{document}